\documentstyle[prl,aps,twocolumn,epsf,floats]{revtex}

\newcommand{\beq}{\begin{equation}}
\newcommand{\eeq}{\end{equation}}
\newcommand{\bea}{\begin{eqnarray}}
\newcommand{\eea}{\end{eqnarray}}

\def\tit#1#2#3#4#5{{#1} {\bf #2}, #3 (#4)}

\def\prl{Phys.\ Rev.\ Lett.\ }
\def\pr{Phys.\ Rev.\ }

\def\state#1{\left|{#1}\right>}

\begin{document}
\draft

\twocolumn[\hsize\textwidth\columnwidth\hsize\csname @twocolumnfalse\endcsname

\title{An RVB phase in the triangular lattice quantum dimer model}
 
\author{R. Moessner and S. L. Sondhi}

\address{Department of Physics, Princeton University,
Princeton, NJ 08544, USA}
\date{\today}

\maketitle

\begin{abstract}

We study the quantum dimer model on the triangular lattice, which
is expected to describe the singlet dynamics of frustrated 
Heisenberg models in phases where valence bond configurations 
dominate their physics. 
We find, in contrast to the square lattice, that there is a truly 
short ranged resonating valence bond (RVB) phase with no gapless collective
excitations and with deconfined, gapped, spinons for a {\it finite}
range of parameters. We also establish the presence of three crystalline
phases in this system.

\end{abstract}

]

The search for an RVB phase of frustrated magnets, inspired by ideas
of Pauling \cite{paulingrvb} and begun in earnest by Anderson 
\cite{Fazekas74}, has been one of the
recurrent themes in research on the cuprate superconductors over the
past decade and somewhat more. Shortly after Anderson's 1987 paper on
the cuprates \cite{pwa87}, RVB theory bifurcated---with a gapless RVB
state (with valence bonds on many length scales) pursued largely by
means of gauge theoretic treatments introduced by Baskaran and
Anderson \cite{ba} defining one track, while Kivelson, Rokhsar and
Sethna \cite{kivrokset} hewed closely to the original vision and
pursued the study of a short ranged RVB state. The latter proposal, of
a state with exponentially decaying spin-spin correlations and no long
range valence bond order, was expected to lead to a gapped collective
excitation spectrum and deconfined, gapped, spinons.  Unfortunately,
it turned out that this was hard to arrange on the square
lattice---the simplest implementation of RVB ideas, the quantum dimer
model \cite{Rokhsar88}, exhibits crystalline order and confined
spinons, except at a critical point \cite{qdmstudy}.  This fact is of a 
piece with an instability of the paramagnetic phase of the $O(3)$ non-linear
sigma model to breaking translational symmetry due to Berry phase effects
\cite{berryphase}.

In this Letter, we report that the simplest quantum dimer model on the
triangular lattice {\it does} possess a short ranged RVB phase with
gapped collective modes, gapped deconfined spinons and spin-charge
separation in its charged excitation spectrum. We establish the presence 
of nearby crystalline phases with confined spinons. We also suggest
that a connection, made previously by P. Chandra and ourselves
\cite{mcs2000}, between quantum dimer models and frustrated transverse
field Ising models can, in this problem, be plausibly extended to track 
the transition out of the RVB phase. This conjecture is closely connected
to the spinon deconfinement mechanism proposed by Read and Sachdev 
\cite{readsach1,sachvojt}, Wen \cite{wen} and the ideas of Senthil and
Fisher \cite{sentfish}.

We study the Rokhsar-Kivelson quantum dimer Hamiltonian generalized to
the triangular lattice (as the contrasts are instructive, we will
comment on the square lattice results along the way):
\bea 
\hat{H} &=& -t \hat{T}+ v \hat{V}
=
\sum_{i=1}^{N_p}\left\{ -t
\sum_{\alpha=1}^3 \left(|\setlength{\unitlength}{3158sp}%
\begingroup\makeatletter\ifx\SetFigFont\undefined%
\gdef\SetFigFont#1#2#3#4#5{%
  \reset@font\fontsize{#1}{#2pt}%
  \fontfamily{#3}\fontseries{#4}\fontshape{#5}%
  \selectfont}%
\fi\endgroup%
\begin{picture}(319,210)(517,-186)
\thinlines
\multiput(720,-10)(1.54799,-2.57998){58}{\makebox(2.0833,14.5833){\SetFigFont{5}{6}{\rmdefault}{\mddefault}{\updefault}.}}
\multiput(718,-10)(-1.55263,-2.58772){58}{\makebox(2.0833,14.5833){\SetFigFont{5}{6}{\rmdefault}{\mddefault}{\updefault}.}}
\multiput(548,-10)(1.55263,-2.58772){58}{\makebox(2.0833,14.5833){\SetFigFont{5}{6}{\rmdefault}{\mddefault}{\updefault}.}}
\multiput(718,-10)(-3.00000,0.00000){58}{\makebox(2.0833,14.5833){\SetFigFont{5}{6}{\rmdefault}{\mddefault}{\updefault}.}}
\multiput(805,-158)(-3.00000,0.00000){58}{\makebox(2.0833,14.5833){\SetFigFont{5}{6}{\rmdefault}{\mddefault}{\updefault}.}}
\thicklines
\put(720, -9){\circle{28}}
\put(803,-153){\line(-1, 0){171}}
\put(547, -9){\circle{28}}
\put(722, -9){\line(-1, 0){172}}
\put(800,-152){\circle{28}}
\put(627,-152){\circle{28}}
\end{picture}
\rangle
\langle \setlength{\unitlength}{3158sp}%
\begingroup\makeatletter\ifx\SetFigFont\undefined%
\gdef\SetFigFont#1#2#3#4#5{%
  \reset@font\fontsize{#1}{#2pt}%
  \fontfamily{#3}\fontseries{#4}\fontshape{#5}%
  \selectfont}%
\fi\endgroup%
\begin{picture}(317,216)(525,-421)
\thinlines
\multiput(720,-235)(1.54799,-2.57998){58}{\makebox(2.0833,14.5833){\SetFigFont{5}{6}{\rmdefault}{\mddefault}{\updefault}.}}
\multiput(718,-235)(-1.55263,-2.58772){58}{\makebox(2.0833,14.5833){\SetFigFont{5}{6}{\rmdefault}{\mddefault}{\updefault}.}}
\multiput(548,-235)(1.55263,-2.58772){58}{\makebox(2.0833,14.5833){\SetFigFont{5}{6}{\rmdefault}{\mddefault}{\updefault}.}}
\multiput(718,-235)(-3.00000,0.00000){58}{\makebox(2.0833,14.5833){\SetFigFont{5}{6}{\rmdefault}{\mddefault}{\updefault}.}}
\multiput(805,-383)(-3.00000,0.00000){58}{\makebox(2.0833,14.5833){\SetFigFont{5}{6}{\rmdefault}{\mddefault}{\updefault}.}}
\thicklines
\put(643,-386){\circle{28}}
\multiput(809,-386)(-5.48713,9.14522){17}{\makebox(8.3333,12.5000){\SetFigFont{7}{8.4}{\rmdefault}{\mddefault}{\updefault}.}}
\put(556,-237){\circle{28}}
\multiput(644,-388)(-5.53125,9.21875){17}{\makebox(8.3333,12.5000){\SetFigFont{7}{8.4}{\rmdefault}{\mddefault}{\updefault}.}}
\put(807,-385){\circle{28}}
\put(720,-234){\circle{28}}
\end{picture}
|+h.c. \right) \right. 
\label{eq:triham}\\
&+& \left. v
\sum_{\alpha=1}^3  \left( |\setlength{\unitlength}{3158sp}%
\begingroup\makeatletter\ifx\SetFigFont\undefined%
\gdef\SetFigFont#1#2#3#4#5{%
  \reset@font\fontsize{#1}{#2pt}%
  \fontfamily{#3}\fontseries{#4}\fontshape{#5}%
  \selectfont}%
\fi\endgroup%
\begin{picture}(319,210)(517,-186)
\thinlines
\multiput(720,-10)(1.54799,-2.57998){58}{\makebox(2.0833,14.5833){\SetFigFont{5}{6}{\rmdefault}{\mddefault}{\updefault}.}}
\multiput(718,-10)(-1.55263,-2.58772){58}{\makebox(2.0833,14.5833){\SetFigFont{5}{6}{\rmdefault}{\mddefault}{\updefault}.}}
\multiput(548,-10)(1.55263,-2.58772){58}{\makebox(2.0833,14.5833){\SetFigFont{5}{6}{\rmdefault}{\mddefault}{\updefault}.}}
\multiput(718,-10)(-3.00000,0.00000){58}{\makebox(2.0833,14.5833){\SetFigFont{5}{6}{\rmdefault}{\mddefault}{\updefault}.}}
\multiput(805,-158)(-3.00000,0.00000){58}{\makebox(2.0833,14.5833){\SetFigFont{5}{6}{\rmdefault}{\mddefault}{\updefault}.}}
\thicklines
\put(720, -9){\circle{28}}
\put(803,-153){\line(-1, 0){171}}
\put(547, -9){\circle{28}}
\put(722, -9){\line(-1, 0){172}}
\put(800,-152){\circle{28}}
\put(627,-152){\circle{28}}
\end{picture}
\rangle
\langle \setlength{\unitlength}{3158sp}%
\begingroup\makeatletter\ifx\SetFigFont\undefined%
\gdef\SetFigFont#1#2#3#4#5{%
  \reset@font\fontsize{#1}{#2pt}%
  \fontfamily{#3}\fontseries{#4}\fontshape{#5}%
  \selectfont}%
\fi\endgroup%
\begin{picture}(319,210)(517,-186)
\thinlines
\multiput(720,-10)(1.54799,-2.57998){58}{\makebox(2.0833,14.5833){\SetFigFont{5}{6}{\rmdefault}{\mddefault}{\updefault}.}}
\multiput(718,-10)(-1.55263,-2.58772){58}{\makebox(2.0833,14.5833){\SetFigFont{5}{6}{\rmdefault}{\mddefault}{\updefault}.}}
\multiput(548,-10)(1.55263,-2.58772){58}{\makebox(2.0833,14.5833){\SetFigFont{5}{6}{\rmdefault}{\mddefault}{\updefault}.}}
\multiput(718,-10)(-3.00000,0.00000){58}{\makebox(2.0833,14.5833){\SetFigFont{5}{6}{\rmdefault}{\mddefault}{\updefault}.}}
\multiput(805,-158)(-3.00000,0.00000){58}{\makebox(2.0833,14.5833){\SetFigFont{5}{6}{\rmdefault}{\mddefault}{\updefault}.}}
\thicklines
\put(720, -9){\circle{28}}
\put(803,-153){\line(-1, 0){171}}
\put(547, -9){\circle{28}}
\put(722, -9){\line(-1, 0){172}}
\put(800,-152){\circle{28}}
\put(627,-152){\circle{28}}
\end{picture}
|+
|\setlength{\unitlength}{3158sp}%
\begingroup\makeatletter\ifx\SetFigFont\undefined%
\gdef\SetFigFont#1#2#3#4#5{%
  \reset@font\fontsize{#1}{#2pt}%
  \fontfamily{#3}\fontseries{#4}\fontshape{#5}%
  \selectfont}%
\fi\endgroup%
\begin{picture}(317,216)(525,-421)
\thinlines
\multiput(720,-235)(1.54799,-2.57998){58}{\makebox(2.0833,14.5833){\SetFigFont{5}{6}{\rmdefault}{\mddefault}{\updefault}.}}
\multiput(718,-235)(-1.55263,-2.58772){58}{\makebox(2.0833,14.5833){\SetFigFont{5}{6}{\rmdefault}{\mddefault}{\updefault}.}}
\multiput(548,-235)(1.55263,-2.58772){58}{\makebox(2.0833,14.5833){\SetFigFont{5}{6}{\rmdefault}{\mddefault}{\updefault}.}}
\multiput(718,-235)(-3.00000,0.00000){58}{\makebox(2.0833,14.5833){\SetFigFont{5}{6}{\rmdefault}{\mddefault}{\updefault}.}}
\multiput(805,-383)(-3.00000,0.00000){58}{\makebox(2.0833,14.5833){\SetFigFont{5}{6}{\rmdefault}{\mddefault}{\updefault}.}}
\thicklines
\put(643,-386){\circle{28}}
\multiput(809,-386)(-5.48713,9.14522){17}{\makebox(8.3333,12.5000){\SetFigFont{7}{8.4}{\rmdefault}{\mddefault}{\updefault}.}}
\put(556,-237){\circle{28}}
\multiput(644,-388)(-5.53125,9.21875){17}{\makebox(8.3333,12.5000){\SetFigFont{7}{8.4}{\rmdefault}{\mddefault}{\updefault}.}}
\put(807,-385){\circle{28}}
\put(720,-234){\circle{28}}
\end{picture}
\rangle
\langle \setlength{\unitlength}{3158sp}%
\begingroup\makeatletter\ifx\SetFigFont\undefined%
\gdef\SetFigFont#1#2#3#4#5{%
  \reset@font\fontsize{#1}{#2pt}%
  \fontfamily{#3}\fontseries{#4}\fontshape{#5}%
  \selectfont}%
\fi\endgroup%
\begin{picture}(317,216)(525,-421)
\thinlines
\multiput(720,-235)(1.54799,-2.57998){58}{\makebox(2.0833,14.5833){\SetFigFont{5}{6}{\rmdefault}{\mddefault}{\updefault}.}}
\multiput(718,-235)(-1.55263,-2.58772){58}{\makebox(2.0833,14.5833){\SetFigFont{5}{6}{\rmdefault}{\mddefault}{\updefault}.}}
\multiput(548,-235)(1.55263,-2.58772){58}{\makebox(2.0833,14.5833){\SetFigFont{5}{6}{\rmdefault}{\mddefault}{\updefault}.}}
\multiput(718,-235)(-3.00000,0.00000){58}{\makebox(2.0833,14.5833){\SetFigFont{5}{6}{\rmdefault}{\mddefault}{\updefault}.}}
\multiput(805,-383)(-3.00000,0.00000){58}{\makebox(2.0833,14.5833){\SetFigFont{5}{6}{\rmdefault}{\mddefault}{\updefault}.}}
\thicklines
\put(643,-386){\circle{28}}
\multiput(809,-386)(-5.48713,9.14522){17}{\makebox(8.3333,12.5000){\SetFigFont{7}{8.4}{\rmdefault}{\mddefault}{\updefault}.}}
\put(556,-237){\circle{28}}
\multiput(644,-388)(-5.53125,9.21875){17}{\makebox(8.3333,12.5000){\SetFigFont{7}{8.4}{\rmdefault}{\mddefault}{\updefault}.}}
\put(807,-385){\circle{28}}
\put(720,-234){\circle{28}}
\end{picture}
|\right)
\right\} \ .\nonumber
\eea Here, the sum on $i$\ runs over all of the $N_p$\ plaquettes, and
the sum on $\alpha$\ over the three different orientations of the
dimer plaquettes, namely \setlength{\unitlength}{3158sp}%
\begingroup\makeatletter\ifx\SetFigFont\undefined%
\gdef\SetFigFont#1#2#3#4#5{%
  \reset@font\fontsize{#1}{#2pt}%
  \fontfamily{#3}\fontseries{#4}\fontshape{#5}%
  \selectfont}%
\fi\endgroup%
\begin{picture}(282,173)(535,-171)
\thinlines
\multiput(720,-10)(1.54799,-2.57998){58}{\makebox(2.0833,14.5833){\SetFigFont{5}{6}{\rmdefault}{\mddefault}{\updefault}.}}
\multiput(805,-158)(-3.00000,0.00000){58}{\makebox(2.0833,14.5833){\SetFigFont{5}{6}{\rmdefault}{\mddefault}{\updefault}.}}
\multiput(718,-10)(-1.55263,-2.58772){58}{\makebox(2.0833,14.5833){\SetFigFont{5}{6}{\rmdefault}{\mddefault}{\updefault}.}}
\multiput(718,-10)(-3.00000,0.00000){58}{\makebox(2.0833,14.5833){\SetFigFont{5}{6}{\rmdefault}{\mddefault}{\updefault}.}}
\multiput(548,-10)(1.55263,-2.58772){58}{\makebox(2.0833,14.5833){\SetFigFont{5}{6}{\rmdefault}{\mddefault}{\updefault}.}}
\end{picture}
 rotated by 0 and
$\pm 60^o$.  We refer to the plaquettes with a parallel pair of dimers 
as flippable plaquettes. As a complete orthonormal basis set we use
$\left\{ \left| c\right> | \ c=1 ... N_c \right\}$, where $\left|
c\right>$\ stands for one of the $N_c$\ possible hardcore dimer
coverings of the triangular lattice. $\hat{V}$\ is diagonal in this
basis, with $\hat{V}\state{c} \equiv n_{fl}(c)\state{c}$ measuring
the number, $n_{fl}(c)$, of flippable plaquettes in configuration
$c$.

Rokhsar and Kivelson \cite{Rokhsar88} derived the square lattice
version of $\hat{H}$ as the leading effective Hamiltonian in 
the singlet manifold consisting of nearest neighbor valence bond
coverings of the lattice by utilizing their overlaps as small parameters; 
subsequently, it was shown by Read and Sachdev \cite{readsach2} that the 
purely kinetic energy ($\hat{T}$) piece described the $1/N$ dynamics of the
nearest-neighbor $SU(N)$ Heisenberg magnet in an extreme quantum limit. 
The former derivation is readily generalized to the 
triangular lattice and, crucially in this case, yields $t>0$ which we
assume in the remainder; the latter is specific to bipartite lattices.

\noindent
{$\mathbf T \gg v,t$}: At high temperatures, but less than the gap
to non-valence bond states, static properties are obtained by a classical
sum over all dimer configurations. Most crisply, consider $T=\infty$
where equal time correlators are given by unweighted averages. The
square lattice problem is {\it critical} in this limit, with algebraically
decaying dimer-dimer correlations \cite{Fisher63}, whence it is not surprising
that at $T=0$ it orders everywhere except at a point. 

The important observation that motivated the present work, is that the triangular
lattice problem is {\it disordered} in this limit with exponentially decaying 
dimer correlations. For example, the correlation function for two parallel dimers
separated by distance $x$ along a column can be computed by standard 
Pfaffian/Grassman methods \cite{fn-triclassical} as,
\bea
\langle n(x) n(0) \rangle &=& ({1\over 6})^2 + G_A^2(x) + G_B(x) G_B(-x) \\
{\rm where}\ \ G_{\{A;B\}}(x) 
&=& \int {d^2k \over (2 \pi)^2 } e^{-ik_x x} {\{2i\sin{k_x};
\bar g({\mathbf k}) \} \over 4 \sin^2k_x + |g({\mathbf k})|^2} \nonumber
\eea
with $g({\mathbf k}) = i[1 - e^{-ik_x} -  e^{-ik_y} - e^{-i(k_x + k_y)}]$. 
It is easy to see that the integrands are analytic in a finite interval about real 
values of the momenta, whence the two Green functions, $G_{\{A;B\}}$, decay 
exponentially with $x$.

This feature {\it already} yields a high temperature RVB phase, and makes a liquid 
phase far more likely at zero temperature, to which we now turn.

\begin{figure}
\epsfxsize=3in
\centerline{\epsffile{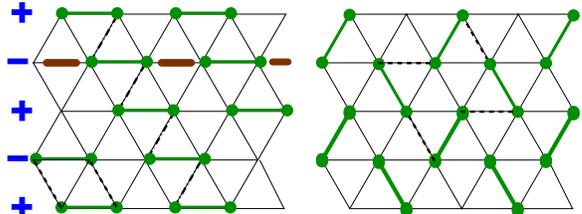}}
\caption{Left: The columnar dimer state. The elementary dimer
plaquette move generated by $\hat{T}$\ is indicated in the bottom left
plaquette. Such plaquette moves conserve the difference between the
number of dimers in rows marked by pluses and minuses.  Dimer moves A
and B, consist of shifting dimers onto the
fat and dot-dashed bonds, respectively. Right: The staggered state
with the four-dimer move connecting it to other states.  }
\label{fig:mfs}
\end{figure}

\noindent
{\bf Topological Sectors:} An important ingredient of the analysis of 
the square lattice problem is the existence of two integer winding 
numbers, for periodic boundary conditions, that are conserved by any 
local dynamics in the dimer
Hilbert space \cite{Rokhsar88}. In that problem it is believed that 
the simplest dynamics (the analog of our $\hat{T}$) is ergodic within
each of the $O(L^2)$\ sectors.
On the triangular lattice, the situation appears to be quite different. 
Repeating the analysis for the square case uncovers only four sectors
corresponding to different combinations of even and odd windings which
will be equivalent in the thermodynamic limit. With the minimal
dynamics of $\hat{T}$ we have found that large classes of configurations
can be connected to each other with no further constraint. The notable
exceptions are the ``staggered'' configuration in Fig.~\ref{fig:mfs}
and its symmetry related counterparts, which have no flippable plaquettes 
whatsoever.
But even here a local four dimer rearrangement can be shown to allow
a connection to other ``generic'' states. We conjecture therefore that
a general local dimer dynamics (which may require as little as the
inclusion of a four dimer move in addition to $\hat{T}$) will be ergodic
in each of the four topological sectors and that the dynamics in our
model is nearly so with the exception of the dynamically disconnected
non-flippable configurations. Note finally, that the Perron-Frobenius 
theorem implies that the ground state of $\hat{H}$ in {\it each} sector
is nodeless.

\begin{figure}
\epsfxsize=3in
\centerline{\epsffile{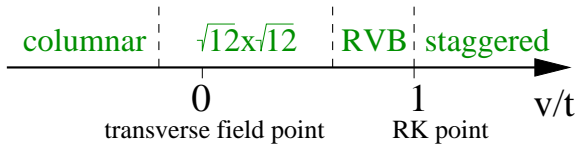}}
\caption{Phase diagram of the quantum dimer model on
the triangular lattice. The nature of the ordered phases is indicated 
above the axis. }
\label{fig:bas}
\end{figure}

\noindent
{\bf RVB Phase, $\mathbf{v_c < v \le t:}$} As in the square lattice
case, the triangular lattice dimer model has a ``Rokhsar-Kivelson''
(RK) point at $v=t$ at which the ground states are equal amplitude 
superpositions of all dimer coverings in a given sector.
To see this, note that a lower bound for the ground state energy is
obtained by considering each plaquette individually. A non-flippable
plaquette is annihilated by $\hat{H}$, whereas a flippable plaquette has a
potential energy of $v$\ and a kinetic energy of, at best, $-t$,
which implies $E_0\geq \min \{0,N_p (v-t)\}$. The equal amplitude
state $\left| RK\right> =N_c^{-1/2}\sum_{c=1}^{N_c}\left| c\right>$
(in any sector) has energy 
$\left< RK \right| H \left| RK\right>=(v-t)\left<n_{fl}\right>$
which vanishes and saturates the lower bound at $v=t$. Following
our statements on the sectoral organization, we conclude that with
the exception of the non-flippable configurations which trivially
saturate this bound, there are four topologically degenerate ground 
states in the thermodynamic limit.
The sum over {\it all} ground states is, for the purposes of
computing correlations diagonal in the dimer basis, equivalent
to the classical dimer problem. As the staggered configurations
are irrelevant to this sum, we conclude that the four generic sector
states have exponentially decaying dimer correlations---i.e. they
are RVB states.

We next wish to argue that these states are representative of a
phase at $T=0$. We will argue shortly that their coexistence with
the staggered states is due to a first order transition
out of the RVB phase exactly at the RK point---much as in the
square lattice problem. Consequently, we wish to establish that
there is a range $v_c < v \le t$ over which the RVB character
of the ground state persists. Typically, one might anticipate
that a disordered ground state goes along with a gap to local
excitations (as opposed to the degeneracy with globally distinct
states). We have examined candidates for collective modes in
the single mode approximation and found that they are all gapped.
We note that, in contrast, Rokhsar and Kivelson found that their 
critical RVB state supported gapless excitations they dubbed resonons
\cite{fn-collec}. These together---the disordered character of the 
ground state and the gap in the local excitation spectrum---are 
strong evidence that the RK point is part of an RVB phase which 
is displaced to its right but will persist to its left (in 
Fig.~\ref{fig:bas}).

To test this argument, we have carried out quantum Monte Carlo
simulations on systems upto $36 \times 36$ sites at temperatures
as low as $t/10$ using the method described in Ref.~\cite{mcs2000}.
As expected, we find that the dimer correlations 
are very short ranged and practically those of the classical dimer
problem, conservatively, for $2/3 < v/t < 1$. We also find
a very weak temperature dependence, suggestive of a gap. 
As $v$ is reduced further, the correlations begin to exhibit crystalline 
structure (see below). At $v=t$ we find a hysteretic (first order)
transition to the staggered phase described next.

\noindent
{\bf Staggered Phase, $\mathbf{v>t}$:} For $v>t$ the lower bound
derived previously implies $E_0\geq 0$. As the staggered 
states are zero energy eigenstates of 
$\hat{H}$ they are the ground states in this range. This
is similar to what happens in the square lattice problem, but
the degeneracy is much lower in our problem ($O(L^0)$ vs $O(L)$) 
and the exactness of the states is here a consequence of dynamics 
rather than topology as mentioned earlier.

An interesting consequence of this last observation is that the
excitation gap is $O(L^0)$ in system size for the staggered phase 
\cite{fn-sqstagg}. At $v/t =\infty$, the lowest energy excitations
are the four dimer loop rearrangement of Fig.~\ref{fig:mfs}. At 
finite but large $v/t$, these get dressed by additional plaquette
flips and acquire a dispersion. This branch of solitons is
expected, by continuity, to be the relevant set of low energy
excitations in the staggered phase even close to the RK point.

\noindent
{\bf Columnar Phase, $\mathbf{-v\gg t:}$} To complete the phase
diagram, we move leftwards from the RVB phase in Fig.~\ref{fig:bas},
turning first to the extreme case $v/t=-\infty$. In this limit,
the kinetic term $\hat{T}$\ is disregarded, and the ground
states are the maximally flippable states, i.e. those states
$\state{c}$\ with maximal $n_{fl}(c)$. To identify the maximally
flippable states, we note that each dimer can be part of at most two
flippable pairs. Since the total number of dimers is fixed, the
maximally flippable states are those in which each dimer belongs to
two pairs.

All maximally flippable states, whose number is exponential
in $L$, can be obtained by carrying out the two operations 
A and B on a particular maximally flippable state as depicted in 
Fig.~\ref{fig:mfs}. 
Pairs of operations of either type as well as global translations
and rotations can be shown to be generated by local plaquette flips. 
This illustrates the point made earlier about winding number
sectors. All maximally flippable states differing by an even number of
operations are in the same sector, whereas a single A or B operation
generates a state in another.

Turning to the case of large but finite $- v/t$, we construct a
perturbation theory using the small parameter $t/v$. Since any 
two maximally flippable states differ in at
least $O(L)$\ dimers, the degenerate perturbation theory is diagonal
at any finite order. The energy shift of a state $\state{c}$\ at order
$2n$\ in perturbation theory depends on the number of flippable
plaquettes, $n_{fl}(c^\prime)$, of the states $\state{c^\prime}$\
which can be reached from $\state{c}$\ by at most $n$\ plaquette
flips. 

The result of this perturbation theory is a striking example of the
phenomenon of quantum ``order by disorder''--- we find that it
selects the columnar state depicted in Fig.~\ref{fig:mfs}. States 
obtained by operations of type A are disfavored at $4^{\rm th}$ order in 
perturbation theory, those generated by type B operations at $6^{\rm th}$
order.

\noindent
{\bf Transverse Field Ising Point, $\mathbf{v=0:}$}  Together
with P. Chandra \cite{mcs2000}, we have recently shown that there
exists an {\em exact} correspondence between the quantum dimer model
at $v=0$, and the fully frustrated transverse field Ising model 
(FFTFIM) on the dual hexagonal lattice at fields $\Gamma$ much
smaller than the magnitude of the exchange $J$ \cite{fn-imdim}. 
In this limit, the
quantum ground state is constructed entirely out of the ground states
of the classical frustrated model and the latter are (upto Ising 
degeneracy), in unique correspondence with dimer coverings of its dual,
triangular, lattice. In our analysis we found a low temperature crystalline
``$\sqrt{12} \times \sqrt{12}$'' phase which exhibits, in dimer language, 
a triangular superlattice with a 12 site unit cell, that is consistent with
a Landau-Ginzburg analysis of the Ising model. While the ordering observed
at accessible system sizes is not conclusive regarding the fate of the 
model at $T=0$, it appears that the columnar phase gives way 
to a different crystalline phase in the proximity of $v=0$.

\noindent
{\bf Spinons:} As noted before, the RVB phase has a gap to collective
excitations. That is also true of the crystalline phases. Also of
interest is the question of confinement for spinons---the gapped spin $1/2$ 
excitations produced by breaking a valence bond. In the dimer
model, these are represented by monomers or holons that carry a spin
and so {\it the questions of holon and spinon confinement are identical}.
This question is most easily addressed by considering the free energies
of states in which two monomers are held a fixed distance apart. At
high temperatures, this is again a classical computation and it is
clear that the spinons are deconfined on the triangular lattice
\cite{rmslsunpub}. The contrast with the square lattice is again 
instructive, for there the spinons are confined at high temperatures.
(However the confinement is very weak, only logarithmic \cite{Fisher63}, which
explains why the more disordered triangular lattice does not confine.)  
At $T=0$ one can readily show
that the state with an equal amplitude sum over dimer configurations with
two spinons localized a fixed distance apart is an eigenstate at the RK
point with an energy {\it independent} of their separation---i.e. the
spinons do not interact beyond one lattice constant \cite{fn-sqspinon}. 
By continuity, we expect spinons to be deconfined in the entire RVB phase
at $T=0$.  From these considerations
it also follows that {\it the charged excitation created by removing an 
electron from the system, will decay into a spinon and holon}. Evidently, 
the holons/spinons will be confined in the crystalline phases.

\noindent
{\bf Spinon Confinement Transition:} Several authors have
suggested that a spinon confinement-deconfinement transition will
be governed by an Ising ($Z_2$) gauge theory \cite{readsach1,wen,sentfish}.
In our own work \cite{mcs2000} we have found that frustrated transverse 
field Ising models (whose duals are precisely the Ising gauge theories)
can provide a description of valence bond phases of Heisenberg
antiferromagnets on their dual lattices via their connection to
quantum dimer models, such as the one considered in this paper.
In the current context, the geometrical identification used by
us leads to an intriguing observation. As already noted, the
quantum dimer model is equivalent to the $\Gamma \ll J$ limit of the 
FFTFIM at $v=0$. Interestingly, the paramagnetic ground state of the 
FFTFIM at $\Gamma \gg J$, {\it when projected onto the dimer manifold}, 
is the equal amplitude RVB sum that is the dimer model ground state at 
$v=t$. This suggests the conjecture that passing between these limits 
in the (projected) FFTFIM gives a description of the spinon confinement 
and translational symmetry breaking transition that takes place at the 
boundary of the RVB and $\sqrt{12} \times \sqrt{12}$ phases. We have 
argued in Ref.~\cite{mcs2000}, that the transition in the unprojected 
model is
in the $O(4)$ universality class. It remains to be seen whether this
identification survives projection. (The remaining transition, between
the columnar and $\sqrt{12} \times \sqrt{12}$ states must be first
order on symmetry grounds.)

In closing, we note that there are two previous ``sightings'' of a spin
liquid phase on the triangular lattice in the literature. First, there
is the large $N$, $Sp(N)$ analysis of Sachdev \cite{sachdev3} which 
found a disordered ground state with gapped unconfined spinons at small 
``spin''. Second, there is the exact diagonalization work of Misguich 
{\it et al.} \cite{misguich} on an $S=1/2$ system with a ferromagnetic 
two spin exchange frustrated by an antiferromagnetic four spin exchange. 
They found a four-fold disordered ground state with a spin gap, but
possibly confined spinons. We hope to clarify the connection between
our results and these in the near future. Finally, we note that along
the lines of the analysis in \cite{Rokhsar88} we expect doping to give
rise, via holon condensation, to a superconducting phase on the triangular
lattice.

As we were finishing this paper, there appeared Ref.~\cite{coldea}, which
reports neutron scattering evidence for deconfined spinons on an anisotropic
triangular lattice. We note that the classical dimer problem in that case
is also disordered and that one can construct an RK point with anisotropic
terms that exhibits a disordered ground state. This suggests that the results
of \cite{coldea} could be understood along the lines of our analysis in this
paper.

\noindent
{\bf Acknowledgements:}
We are grateful to S. Kivelson and D. Huse for very valuable
discussions and to P. Chandra for collaboration on related work. 
This work was supported in part by grants from the Deutsche
Forschungsgemeinschaft, the NSF (grant No. DMR-9978074), the A. P. Sloan
Foundation and the David and Lucille Packard Foundation.


\begin{references}

\bibitem{paulingrvb} L. Pauling, Proc. Nat. Acad. Sci.  {\bf 39}, 
551 (1953).

\bibitem{Fazekas74}
P. W. Anderson, Mat. Res. Bull. {\bf 8}, 153 (1973);
P. Fazekas and P. W. Anderson, {\sl Phil. Mag.\ }{\bf 30}, 23 (1974).

\bibitem{pwa87}
P. W. Anderson, \tit{Science}{235}{1196}{1987}{THE RESONATING VALENCE
BOND STATE IN LA2CUO4 AND SUPERCONDUCTIVITY}.


\bibitem{ba}
G. Baskaran and P. W. Anderson, Phys. Rev. B {\bf 37}, 580 (1988);
P. A. Lee, Physica C {\bf 317-318}, 194 (1999) and references therein.

\bibitem{kivrokset}
S. A. Kivelson, D. S. Rokhsar and J. P. Sethna, Phys. Rev. B {\bf 35},
8865 (1987).

\bibitem{Rokhsar88}
D. S. Rokhsar and S. A. Kivelson,
\tit{\prl}{61}{2376}{1998}{SUPERCONDUCTIVITY AND THE QUANTUM HARD-CORE
DIMER GAS}.

\bibitem{qdmstudy} See e.g. E. Fradkin, {\it Field Theories
of Condensed Matter Systems}, Addison-Wesley (1991). 
A more complete treatment is C. L. Henley (unpublished).

\bibitem{berryphase} F. D. M. Haldane, Phys. Rev. Lett. {\bf 61}, 1029
(1988); N. Read and S. Sachdev, Phys.  Rev. Lett. {\bf 62}, 1694 (1989).

\bibitem{mcs2000} 
R. Moessner, S. L. Sondhi and P. Chandra,
\tit{\prl}{84}{4457}{2000}{Two-dimensional periodic frustrated Ising
models in a transverse field}.

\bibitem{readsach1}
N. Read and S. Sachdev, Phys. Rev. Lett. {\bf 66}, 1773 (1991).

\bibitem{sachvojt}  S. Sachdev and M. Vojta, cond-mat/9910231.

\bibitem{wen} X.G. Wen, Phys. Rev. {\bf B44}, 2664 (1991).

\bibitem{sentfish}
T. Senthil and M. P. A. Fisher, cond-mat/9912380.

\bibitem{readsach2}
N. Read and S. Sachdev, Nucl. Phys. B {\bf 316}, 609 (1989).

\bibitem{Fisher63}
M.E. Fisher and J. Stephenson, \tit{\pr}{132}{1411}{1963}{Statistical 
Mechanics of Dimers on a Plane Lattice. II. Dimer 
Correlations and Monomers}.

\bibitem{fn-triclassical} See, {\it e.g.}, S. Samuel, J. Math. Phys. {\bf 21},
2806 (1980). Our results for the triangular lattice have not, to our knowledge,
appeared previously in the literature; we expect to report more details 
elsewhere.

\bibitem{fn-collec} Gaplessness on the square lattice 
relies on two ingredients in a single-mode calculation: 
a conservation law, which is present in our case too, 
for the Hamiltonian only creates and destroys pairs of parallel 
neighboring dimers and that the conserved quantity, defined in 
Fig.~\ref{fig:mfs}, has an extensive expectation value 
in a ground state. The conserved quantity is always subextensive
on the triangular lattice, at most $O(L)$, and the resonons are gapped

\bibitem{fn-sqstagg}
The corresponding gap in the square lattice staggered phase 
is $O(L)$, which therefore has {\it no} local excitations 
in the thermodynamic limit.

\bibitem{fn-imdim} There are similar mappings for the square and
hexagonal lattice dimer models \cite{mcs2000}.

\bibitem{rmslsunpub} R. Moessner and S. L. Sondhi (unpublished).

\bibitem{fn-sqspinon} This is also true at the RK point on the square 
lattice and is implicit in the results of \cite{Rokhsar88}---note
the remarkable feature that raising the temperature {\it confines}
the spinons. Elsewhere in the square lattice phase diagram, the spinons 
are confined.

\bibitem{sachdev3} S. Sachdev, Phys. Rev. B {\bf 45}, 12377 (1992).

\bibitem{misguich} G. Misguich {\it et al.}, Phys. Rev. B 60, 1064 (1999).

\bibitem{coldea} R. Coldea {\it et al.}, cond-mat/0007172.


\end{references}
\end{document}